\documentclass[sn-aps]{sn-jnl}

\usepackage{hyperref}
\hypersetup{linkcolor=blue,citecolor=red}

\usepackage{natbib}
\usepackage{amssymb,amsmath}

\usepackage{epsfig}
\usepackage{bm}
\usepackage{color}
\usepackage{graphicx}

\newcommand\beq{\begin{equation}}
\newcommand\eeq{\end{equation}}
\newcommand\beqa{\begin{eqnarray}}
\newcommand\eeqa{\end{eqnarray}}
\def\bal#1\eal{\begin{align}#1\end{align}}
\newcommand{\nn}{\nonumber\\}

\newcommand{\gd}{\dot{\gamma}}
\newcommand{\gda}{\dot{\gamma}^*}
\newcommand{\bw}{\boldsymbol{\omega}}
\newcommand{\bc}{\mathbf{v}}
\newcommand{\bV}{\mathbf{V}}
\newcommand{\ka}{\kappa}
\newcommand{\bt}{\widetilde\beta}
\newcommand{\at}{\widetilde\alpha}
\newcommand{\bg}{\mathbf{g}}
\newcommand{\bk}{\hnn{\sigma}}
\newcommand{\medio}[1]{\left\langle #1\right\rangle}
\newcommand{\al}{\alpha}
\newcommand{\be}{\beta}

\newcommand{\bW}{\boldsymbol{\Omega}}

\newcommand{\bJ}{\mathbf{Q}}

\newcommand{\nuM}{\nu}
\newcommand{\hnn}[1]{\widehat{\boldsymbol #1}}

        \jyear{2026}%

        \theoremstyle{thmstyleone}%
        \theoremstyle{thmstyletwo}%

        \theoremstyle{thmstylethree}%

\begin{document}

\title[Rheology of Uniform Shear Flow]{Exact Rheology of Uniform Shear Flow in a Gas of Inelastic and Rough Maxwell Particles}


\author*[1]{\fnm{Andr\'es} \sur{Santos}}\email{andres@unex.es}
\author[2]{\fnm{Gilberto M.} \sur{Kremer}}\email{kremer@fisica.ufpr.br}

\affil[1]{\orgdiv{Departamento de F\'{\i}sica and Instituto de Computaci\'on Cient\'ifica Avanzada (ICCAEx)}, \orgname{Universidad de Extremadura},
\orgaddress{
\city{Badajoz}, \postcode{E-06006},
\country{Spain}}}

\affil[2]{\orgdiv{Departamento de F\'{\i}sica}, \orgname{Universidade Federal do Paran\'a},
\orgaddress{
\city{Curitiba},
\country{Brazil}}}



\abstract{
We investigate the steady uniform shear flow of a granular gas composed of inelastic and rough Maxwell particles. Exploiting the mean-field character of the model, we derive exact expressions for the collisional production rates of the second-degree moments and obtain a closed nonlinear solution for the stress and spin-spin tensors. The rotational-to-translational temperature ratio and the proportionality between the spin-spin and stress tensors are shown to be independent of the coefficient of normal restitution and determined solely by roughness and moment of inertia. The reduced normal stresses, shear stress, and shear rate are obtained explicitly in terms of two effective parameters generalizing the cooling and stress relaxation rates of the smooth model. From these results we derive exact expressions for the non-Newtonian shear viscosity, the first viscometric function, and the friction coefficient. The dependence of the rheological properties on the normal and tangential restitution coefficients is analyzed in detail, revealing strong non-Newtonian behavior and nonmonotonic effects of roughness. The results reduce, in the appropriate limits, to those of the inelastic Maxwell model for smooth particles and to the Pidduck gas in the elastic perfectly rough case.
}




\maketitle

\section{Introduction}\label{sec1}

The most widely used model for a granular gas is the inelastic hard-sphere model (IHSM), where the grains are assumed to be perfectly smooth spheres colliding with a constant coefficient of normal restitution $\alpha$ \cite{BP04,G19}. A much more tractable alternative is the inelastic Maxwell model (IMM), in which the velocity-dependent collision rate is replaced by an effective mean-field constant \cite{BCG00}. This simplification has been widely exploited to obtain a variety of exact results within the IMM framework. However, both the IHSM and the IMM neglect the impact of surface roughness on the dynamical properties of granular gases. This limitation is overcome by the inelastic rough hard-sphere model (IRHSM), where, in addition to $\alpha$, a constant coefficient of tangential restitution $\beta$ is introduced \cite{JR85a}.

In parallel with the simplification leading from the IHSM to the IMM, we have recently proposed the inelastic rough Maxwell model (IRMM) as a Maxwellian counterpart of the IRHSM.
Again, the essential simplification  is
that the collision rate is replaced by an effective mean-field
frequency, independent of the relative velocity. This property allows
the collisional moments to be expressed exactly in terms of velocity
moments of equal or lower degree.
In a first work, we derived the exact expressions for the most relevant collisional moments \cite{KS22}, and in a second one we obtained the exact Navier--Stokes--Fourier (NSF) transport coefficients \cite{SK24}.

It is well known that nonequilibrium steady states of granular gases are inherently non-Newtonian, so that the stress tensor $\Pi_{ij}$ may significantly deviate from its NSF form \cite{SGD04,G19}. From a theoretical viewpoint, the paradigmatic nonequilibrium state is the uniform (or simple) shear flow (USF). This state is characterized by a constant shear rate $\dot\gamma$ and a spatially uniform velocity distribution function in the Lagrangian frame, $f(\bV)$, where $\bV=\mathbf{v}-\dot\gamma y\mathbf{e}_x$ is the peculiar velocity.

The USF for the IHSM has been extensively studied by means of computer simulations \cite{CG86,C89,C97,MGAL06,CVG15,HTG17,HT19}, kinetic models \cite{BRM97,G06,GG19}, and approximate solutions of the Boltzmann and Enskog equations, such as Grad's moment method \cite{SGD04,AS05,CVG15,HTG17,HT19,THSG20}. In the case of the IMM, the velocity-independence of the collision rate allows for the derivation of exact results \cite{C01b,G03a,G06,G07,SG07,G08c,GT10,GS11,GT12,KGS14,GT15,GKT15,GT16,GG19,KG20,RG23}.

The inclusion of roughness in the IRHSM renders the model more realistic but also considerably increases its mathematical complexity. In analogy with the IHSM, the USF for the IRHSM has been investigated by computer simulations \cite{GA08}, kinetic modeling \cite{S11a,GG20}, and Grad's moment method \cite{GG20}.

The aim of the present work is to complement the previous studies on the non-Newtonian rheological properties of the USF by deriving exact results within the framework of the IRMM.
Whereas our previous study of the NSF transport coefficients~\cite{SK24} addressed
the weak-gradient hydrodynamic regime, the USF constitutes an inherently
non-Newtonian state with finite shear rates~\cite{SGD04}. Therefore, the rheological
properties derived here go beyond the NSF description and provide an
exact characterization of the nonlinear response of the IRMM.

The paper is organized as follows. In Sect.~\ref{sec2} we formulate the Boltzmann equation for the steady uniform shear flow and derive the hierarchy of balance equations for the relevant velocity moments. Section~\ref{sec3} recalls the definition of the IRMM and the exact expressions of the collisional production rates needed for the USF problem. In Sect.~\ref{sec4} we obtain the exact solution for the reduced stress and spin-spin tensors, analyze the resulting nonlinear rheological properties, and discuss their dependence on the coefficients of restitution and the moment of inertia. The main conclusions are summarized in Sect.~\ref{sec5}.

\section{Boltzmann equation for the uniform shear flow}
\label{sec2}
\subsection{Collision rules}

Let us consider a granular gas composed of spherical particles of diameter $\sigma$, mass $m$, and moment of inertia $I$. The translational and angular velocities of a particle are denoted by $\bc$ and $\bw$, respectively.

In both the IRHSM and the IRMM, collisions are characterized by a constant coefficient of normal restitution $\alpha$ and a constant coefficient of tangential restitution $\beta$. The coefficient of normal restitution is positive,  its maximum value ($\alpha=1$) corresponding to elastic collisions. The tangential restitution satisfies the physical bound
$-1\le \beta \le 1$, where the limiting values $\beta=-1$ and
$\beta=1$ correspond to perfectly smooth and perfectly rough particles,
respectively.

If $(\bc_1,\bw_1;\bc_2,\bw_2)$ are the precollisional velocities of a colliding pair, the corresponding postcollisional velocities are given by~\cite{SKG10,G19}
  \beq
\label{1}
\hat{b}_{\bk} \begin{Bmatrix}\bc_1\\\bc_2\end{Bmatrix}=\begin{Bmatrix}\bc_1\\\bc_2\end{Bmatrix} \mp\frac{\bJ}{m},\quad \hat{b}_{\bk}\begin{Bmatrix}\bw_1\\\bw_2\end{Bmatrix}=\begin{Bmatrix}\bw_1\\\bw_2\end{Bmatrix} -{\sigma\over2}\,\bk\times\frac{\bJ}{I},
  \eeq
where $\hat{b}_{\bk}$ is the collision operator in velocity space, $\bk$ is the unit vector joining the centers of the two colliding particles at contact, and $\bJ$ is the impulse exerted by particle $1$ on particle $2$. Its explicit expression is
 \beq
\bJ={m\widetilde\alpha}(\bk\cdot\bg)\bk-m\widetilde\beta\bk\times\left(\bk\times\bg+\sigma\frac{\bw_1+\bw_2}{2}\right),
\label{6J}
 \eeq
where $\bg=\bc_1-\bc_2$ is the relative translational velocity of the centers of mass. The parameters
 \beq\label{7}
 \widetilde\alpha\equiv{1+\alpha\over2},\quad \widetilde\beta\equiv\frac{1+\beta}{2}\frac{\kappa}{1+\kappa},\quad \kappa\equiv \frac{I}{m\sigma^2},
 \eeq
are functions of the coefficients of restitution and of the  moment of inertia.

\subsection{Boltzmann equation}

In the special case of the steady-state USF, the Boltzmann equation takes the form
\beq
\label{USF}
-\gd V_y\frac{\partial f(\bV,\bw)}{\partial V_x}=J[\mathbf{V},\bm{\omega}\vert f,f],
\eeq
where $\dot{\gamma}$ is the constant shear rate and $f(\bV,\bw)$ is the spatially uniform velocity distribution function in the Lagrangian frame.
The left-hand side of Eq.~\eqref{USF} accounts for the effect of the shear field in the co-moving frame, while the right-hand side represents the Boltzmann bilinear collision operator.

Given an arbitrary function $\Psi(\bV,\bw)$, we define its average value $\langle\Psi\rangle$ and its collisional production rate $\mathcal{J}[\Psi]$ as
\begin{subequations}
\beq
\langle\Psi\rangle =\frac{1}{n}\int d\bV\int d\bw\, \Psi(\bV,\bw)f(\bV,\bw),
\label{avpsi}
\eeq
\beq
\label{16J}
\mathcal{J}[\Psi]=\frac{1}{n}\int d\bV\int d\bw\,\Psi(\bV,\bw)J[\bV,\bw\vert f,f],
\eeq
\end{subequations}
where $n$ is the number density. By definition and conservation of linear momentum, one has $\langle\bV\rangle=\mathcal{J}[\bV]=0$.

In the USF problem, the most relevant quantities are the first- and second-degree velocity moments:
\begin{subequations}
\label{4a-4e}
\beq
\bW=\medio{\bw},\quad \Upsilon_{ij}=n\medio{\omega_iV_j},
\eeq
\beq
 T_t=\frac{m}{3}\medio{V^2},\quad T_r=\frac{I}{3}\medio{\omega^2},
 \eeq
\beq
  \Pi_{ij}=n m\medio{V_iV_j-\frac{1}{3}V^2\delta_{ij}}, \quad \Omega_{ij}=nI\medio{\omega_i\omega_j-\frac{1}{3}\omega^2\delta_{ij}}.
\eeq
\end{subequations}
Here, $\bW$ is the mean spin vector, $\Upsilon_{ij}$ is the couple stress tensor~\cite{B97,MHN02}, $T_t$ and $T_r$ are the translational and rotational granular temperatures, respectively, $\Pi_{ij}$ is the (traceless) stress tensor, and $\Omega_{ij}$ is the (traceless) spin-spin tensor. Note that the pressure tensor is
\beq
\label{Pij}
P_{ij}=nm\medio{V_iV_j}=\Pi_{ij}+nT_t\delta_{ij}.
\eeq

Taking moments on both sides of Eq.~\eqref{USF}, one straightforwardly obtains the following balance equations:
\begin{subequations}
\label{8}
\beq
\label{8a}
\mathcal{J}[\bw]=0,\quad n\mathcal{J}[\omega_iV_j]=\gd\Upsilon_{iy}\delta_{jx},
\eeq
\beq
\label{8b}
nm\mathcal{J}[V^2]=2\gd\Pi_{xy},\quad \mathcal{J}[\omega^2]=0,\quad \mathcal{J}\left[\omega_i\omega_j-\frac{1}{3}\omega^2\delta_{ij}\right]=0,
\eeq
\bal
\label{8c}
nm\mathcal{J}\left[V_iV_j-\frac{1}{3}V^2\delta_{ij}\right]=&\gd\bigg[\Pi_{iy}\delta_{jx}+\Pi_{jy}\delta_{ix}-\frac{2}{3}\Pi_{xy}\delta_{ij}
+nT_t\left(\delta_{ix}\delta_{jy}+\delta_{iy}\delta_{jx}\right)\bigg].
\eal
\end{subequations}

Within the IRHSM, the production rates $\mathcal{J}[\Psi]$ cannot, in general, be evaluated exactly. This limitation is overcome in the IRMM, where the mean-field character of the collision rate allows one to obtain closed expressions for those collisional moments.

\section{The inelastic and rough Maxwell model}
\label{sec3}
In the IRMM, the Boltzmann collision operator is given by~\cite{KS22,SK24}
\beq
J[\bV_1,\bw_1\vert f,f]=\frac{\nuM}{4\pi n}\int d\bV_2\int d\bw_2\int d\bk\,\left(\frac{\hat{b}_{\bk}^{-1}}{\alpha\beta^2} -1\right)f(\bV_1,\bw_1)f(\bV_2,\bw_2),
\label{15M}
 \eeq
where $\nuM\propto \sqrt{T_t}$ is an effective mean-field collision frequency.
Unlike the collision frequency of hard-sphere models, the effective
collision frequency $\nuM$ is independent of the relative velocity of
the colliding pair. Nevertheless, it is not a constant, since it depends on the instantaneous
translational temperature.

The mean-field structure of Eq.~\eqref{15M} allows one to evaluate exactly the production rates appearing in Eqs.~\eqref{8}~\cite{KS22}. In particular,
\beq
\label{1a}
 -\nuM^{-1}{\mathcal{J}\left[\bw\right]}= \varphi_{01\mid  01}  \bW,\quad
-\nuM^{-1}n{\mathcal{J}\left[\omega_i V_j\right]}= \psi_{11\mid  11} \Upsilon_{ij},
\eeq
where the coefficients $\varphi_{01\mid 01}$ and $\psi_{11\mid 11}$ depend on $\alpha$, $\beta$, and $\kappa$, although their explicit expressions are not required here. Substitution of Eq.~\eqref{1a} into Eq.~\eqref{8a} implies that $\bW=\Upsilon_{ij}=0$ in the USF state.

Under this condition, the production rates associated with Eqs.~\eqref{8b} and \eqref{8c} become
\begin{subequations}
\label{6A-6H}
\beq
\label{6b}
-\nuM^{-1}\frac{m}{3}{\mathcal{J}\left[V^2\right]}=\chi_{20\mid  20}T_t  +\frac{4}{\ka}\chi_{20\mid  02}T_r,
\eeq
\beq
\label{6c}
-\nuM^{-1}\frac{I}{3}{\mathcal{J}\left[\omega^2\right]}=
  \frac{\ka}{4}\chi_{02\mid  20}T_t+\chi_{02\mid  02}T_r,
\eeq
\beq
\label{6d}
-\nuM^{-1}nm{\mathcal{J}\left[V_iV_j-\frac{1}{3}V^2\delta_{ij}\right]}=
\psi_{20\mid  20}\Pi_{ij}  +\frac{4}{\ka}\psi_{20\mid  02}\Omega_{ij},
\eeq
\beq
-\nuM^{-1}nI{\mathcal{J}\left[\omega_i\omega_j-\frac{1}{3}\omega^2\delta_{ij}\right]}=
 \frac{\ka }{4}\psi_{02\mid  20}\Pi_{ij}+\psi_{02\mid  02}\Omega_{ij}   .
\eeq
\end{subequations}

The explicit expressions of the eight coefficients appearing in Eqs.~\eqref{6A-6H} are~\cite{KS22}
\begin{subequations}
\beq
\label{B2}
\chi_{20\mid 20}=\frac{2}{3}\left[\at\left(1-\at\right)+2\bt\left(1-\bt\right)\right],\quad \chi_{20\mid 02}=-\frac{\bt^2}{3},
\eeq
\beq
\label{B4}
\chi_{02\mid 20}=-\frac{16\bt^2}{3\ka^2},\quad \chi_{02\mid 02}=\frac{4\bt}{3\ka}\left(1-\frac{\bt}{\ka}\right),
\eeq
\beq
\label{B3}
\psi_{20\mid 20}=\frac{2}{15}\left(5\at-2\at^2-6\at\bt+10\bt-7\bt^2\right),
\eeq
\beq
\psi_{20\mid 02}=\frac{\bt^2}{6}, \quad \psi_{02\mid 20}=\frac{8\bt^2}{3\ka^2},\quad \psi_{02\mid 02}=\frac{2\bt}{15\ka}\left(10-\frac{7\bt}{\ka}\right).
\eeq
 \end{subequations}

\section{Results}
\label{sec4}
Insertion of Eqs.~\eqref{6A-6H} into Eqs.~\eqref{8b} and \eqref{8c} yields
\begin{subequations}
\beq
\label{13a}
\theta\equiv \frac{T_r}{T_t}=-\frac{\ka}{4}\frac{\chi_{02\mid 20}}{\chi_{02\mid 02}}=\frac{\ka(1+\be)}{1-\be+2\ka},
\eeq
\beq
\label{6}
\chi_{20\mid 20}+\frac{4}{\ka}\chi_{20\mid 02}\theta=-\frac{2}{3}\gda \Pi_{xy}^*,
\eeq
\beq
\label{8psi}
\psi_{20\mid 20}\Pi_{ij}^*+\frac{4}{\ka}\psi_{20\mid 02}\Omega_{ij}^*={\gda} \left(\frac{2}{3}\Pi_{xy}^*\delta_{ij}-\Pi_{iy}^*\delta_{jx}-
\Pi_{jy}^*\delta_{ix}
-\delta_{ix}\delta_{jy}-\delta_{jx}\delta_{iy}\right),
\eeq
 \beq
 \label{13d}
 \Omega_{ij}^*=-\lambda\Pi_{ij}^*,\quad \lambda=\frac{\ka}{4}\frac{\psi_{02\mid 20}}{\psi_{02\mid 02}}=\frac{5 \ka(1+\be)}{13-7\be+20\ka},
\eeq
\end{subequations}
where $\gda\equiv\gd/\nuM$ is the reduced shear rate, $\Pi_{ij}^*\equiv \Pi_{ij}/nT_t$ is the reduced stress tensor, and $\Omega_{ij}^*\equiv \Omega_{ij}/nT_t$ is the reduced spin-spin tensor.

\begin{figure}[tbp]
\centering
\includegraphics[width=0.5\textwidth]{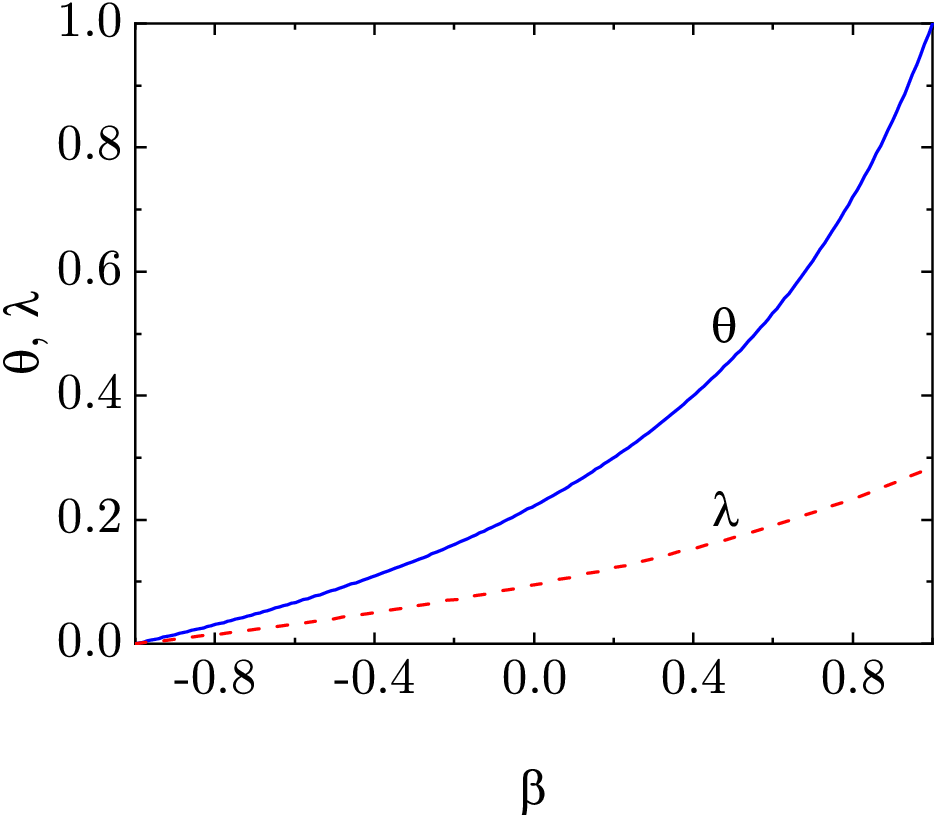}
\caption{
Rotational-to-translational temperature ratio $\theta=T_r/T_t$  and proportionality factor $\lambda$ relating the spin-spin tensor to the stress tensor, $\Omega_{ij}^*=-\lambda \Pi_{ij}^*$, as functions of the tangential restitution coefficient $\beta$ for $\kappa=\frac{2}{5}$.}\label{fig1}
\end{figure}

Equation~\eqref{13a} shows that the rotational-to-translational temperature ratio $\theta$ is independent of the coefficient of normal restitution $\alpha$. It ranges from $\theta=0$ in the perfectly smooth case ($\beta=-1$) to $\theta=1$ in the perfectly rough case ($\beta=1$). This temperature ratio coincides with that obtained from a Bhatnagar--Gross--Krook-like model~\cite{S11a} and also with the one corresponding to the homogeneous steady state driven by a white-noise thermostat.
Moreover, according to Eq.~\eqref{13d}, the spin-spin tensor is proportional to the stress tensor, $\Omega_{ij}^*=-\lambda \Pi_{ij}^*$, with a negative proportionality factor $-\lambda$ that is independent of $\alpha$ and vanishes in the perfectly smooth limit ($\beta=-1$).
Figure~\ref{fig1} displays the $\beta$-dependence of both $\theta$ and $\lambda$ for $\kappa=\frac{2}{5}$.

Substitution of Eq.~\eqref{13a} into Eq.~\eqref{6} gives
 \beq
\label{10}
\frac{2}{3}\gda \Pi_{xy}^*=-\chi,
\eeq
where
\beq
\label{12}
\chi\equiv \chi_{20\mid 20}-\frac{\chi_{20\mid 02}\chi_{02\mid 20}}{\chi_{02\mid 02}}=\frac{2}{3}\left(\frac{1-\al^2}{4}+\ka\frac{1-\be^2}{1-\be+2\ka}\right)
\eeq
is proportional to the translational cooling rate.
Likewise, insertion of Eq.~\eqref{13d} into Eq.~\eqref{8psi} yields
\beq
\label{11}
{\gda}\left(\frac{2}{3}\Pi_{xy}^*\delta_{ij}-\Pi_{iy}^*\delta_{jx}-\Pi_{jy}^*\delta_{ix}-\delta_{ix}\delta_{jy}-\delta_{jx}\delta_{iy}\right)=\psi\Pi_{ij}^*,
\eeq
where
\bal
\psi\equiv \psi_{20\mid 20}-\frac{\psi_{20\mid 02}\psi_{02\mid 20}}{\psi_{02\mid 02}}=&\frac{1+\al}{3}-\frac{(1+\al)^2}{15}+\frac{\ka(1+\be)}{1+\ka}\bigg[\frac{2}{3}-\frac{1+\al}{5}\nn
&-
\frac{7\ka(1+\be)}{30(1+\ka)}-\frac{5\ka(1+\be)^2}{6(1+\ka)(13-7\be+20\ka)}\bigg]
\eal
can be interpreted as the stress relaxation rate.

The structure of Eqs.\ \eqref{10} and \eqref{11} is the same as in the smooth case~\cite{SG07}. The solution  is $\Pi_{xz}^*=\Pi_{yz}^*=0$ and
\begin{subequations}
\beq
\Pi_{xx}^*=-2\Pi_{yy}^*=-2\Pi_{zz}^*=2\frac{\chi}{\psi},
\label{13}
\eeq
\beq
\label{18b}
\Pi_{xy}^*=-\sqrt{\frac{3}{2}\frac{\chi}{\psi}\left(1-\frac{\chi}{\psi}\right)},
\eeq
\beq
\label{18c}
\gda=\sqrt{\frac{3}{2}\frac{\psi\chi}{1-\chi/\psi}}.
\eeq
\end{subequations}
Equation~\eqref{18c} shows that the steady translational temperature scales quadratically with the shear rate. More explicitly,
\beq
T_t=T_t^0\left(\frac{\dot{\gamma}}{\nu^0}\right)^2
\frac{2}{3}\frac{1-\chi/\psi}{\psi\chi},
\eeq
where $T_t^0$ is an arbitrary reference temperature and $\nu^0\propto \sqrt{T_t^0}$ is the associated reference collision frequency.
Therefore, the steady  USF exists only when the translational temperature
takes the value given above, so that viscous heating exactly balances
collisional cooling. In this sense, the reduced shear rate
$\dot{\gamma}^*$ is not an externally prescribed parameter but is
determined self-consistently by the steady-state condition.

Combining Eqs.~\eqref{13} and \eqref{18b}, one obtains the nonequilibrium ``equation of state''
\beq
\Pi_{xy}^*=-\frac{1}{2}\sqrt{\frac{3}{2}\Pi_{xx}^*\left(2-\Pi_{xx}^*\right)},
\eeq
which holds for arbitrary $\alpha$, $\beta$, and $\kappa$, and coincides with that derived from a Bhatnagar--Gross--Krook-like model~\cite{S11a}. Analogously,
\beq
\label{20}
\frac{\gda}{\psi}=\sqrt{\frac{3}{2}\frac{\Pi_{xx}^*}{2-\Pi_{xx}^*}}.
\eeq

\begin{figure}[tbp]
\centering
\includegraphics[height=0.32\textwidth]{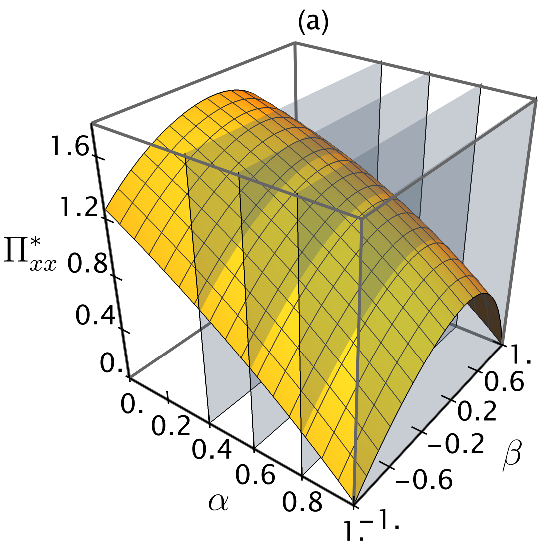}\includegraphics[height=0.32\textwidth]{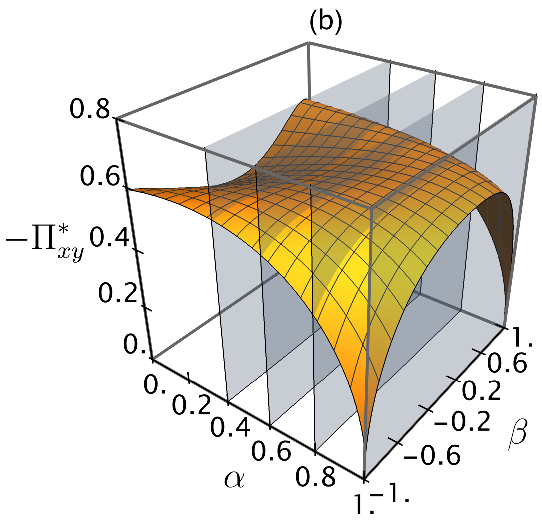}\includegraphics[height=0.32\textwidth]{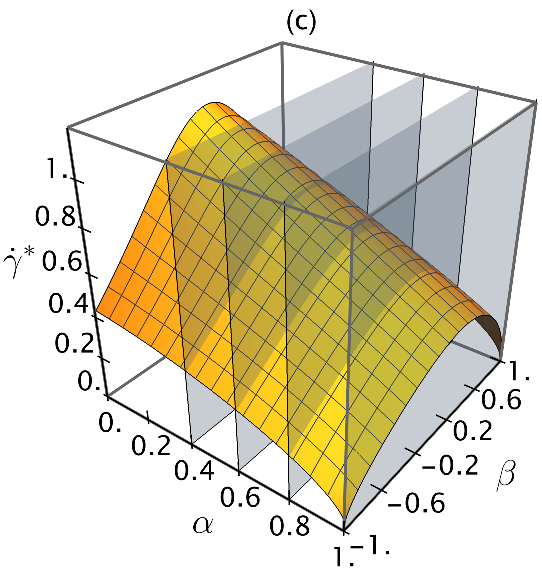}\\
\includegraphics[height=0.28\textwidth]{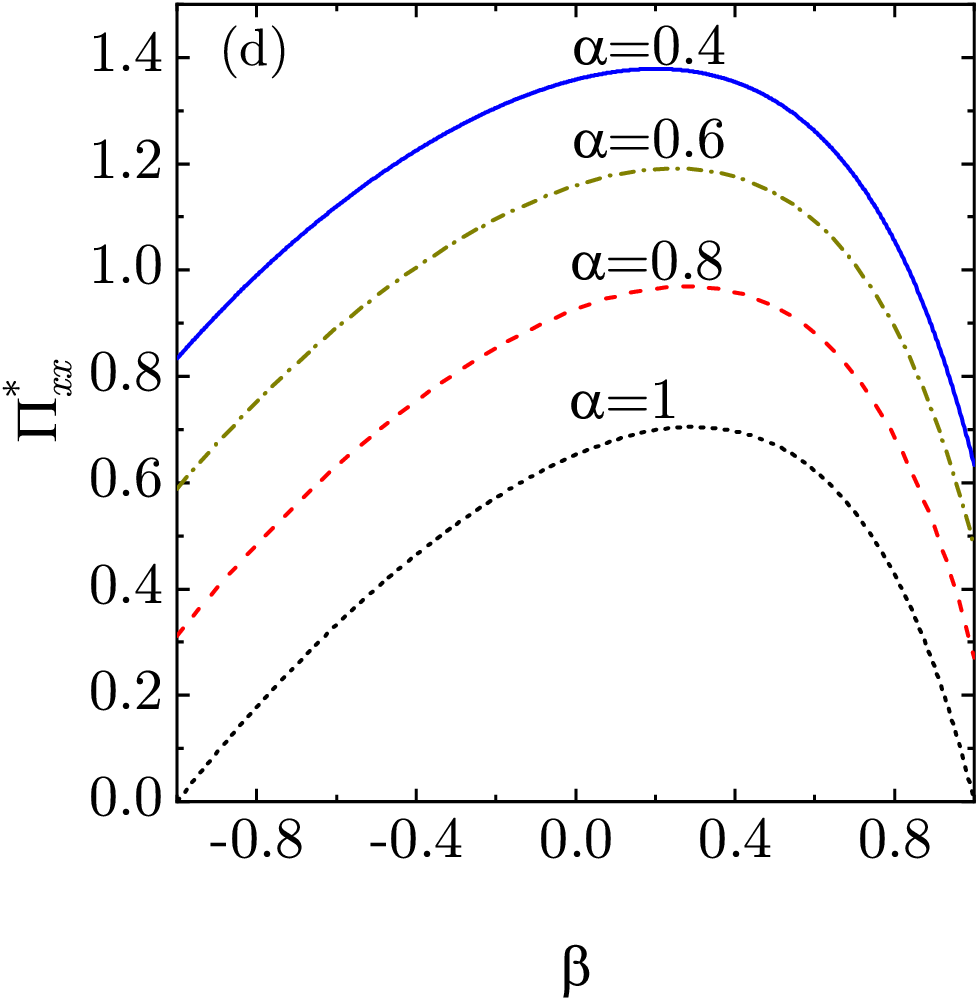}\hspace{0.5cm}\includegraphics[height=0.28\textwidth]{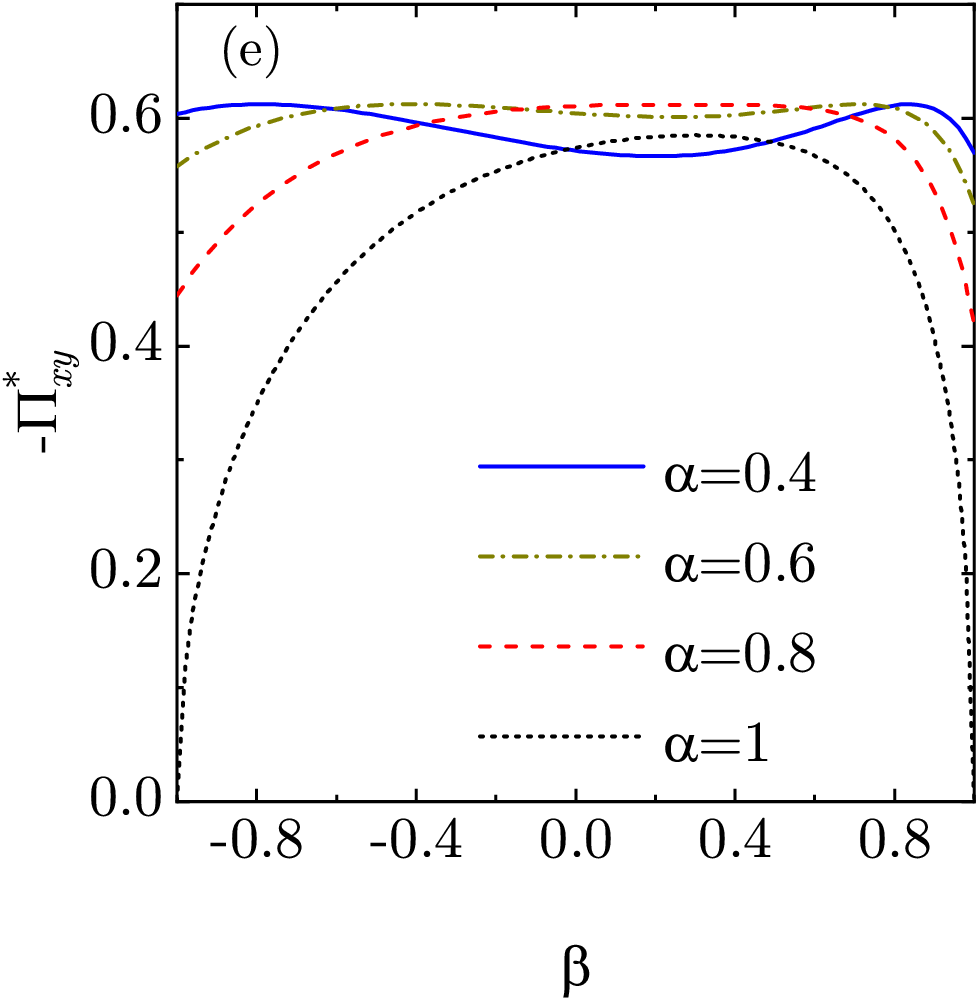}\hspace{0.5cm}\includegraphics[height=0.28\textwidth]{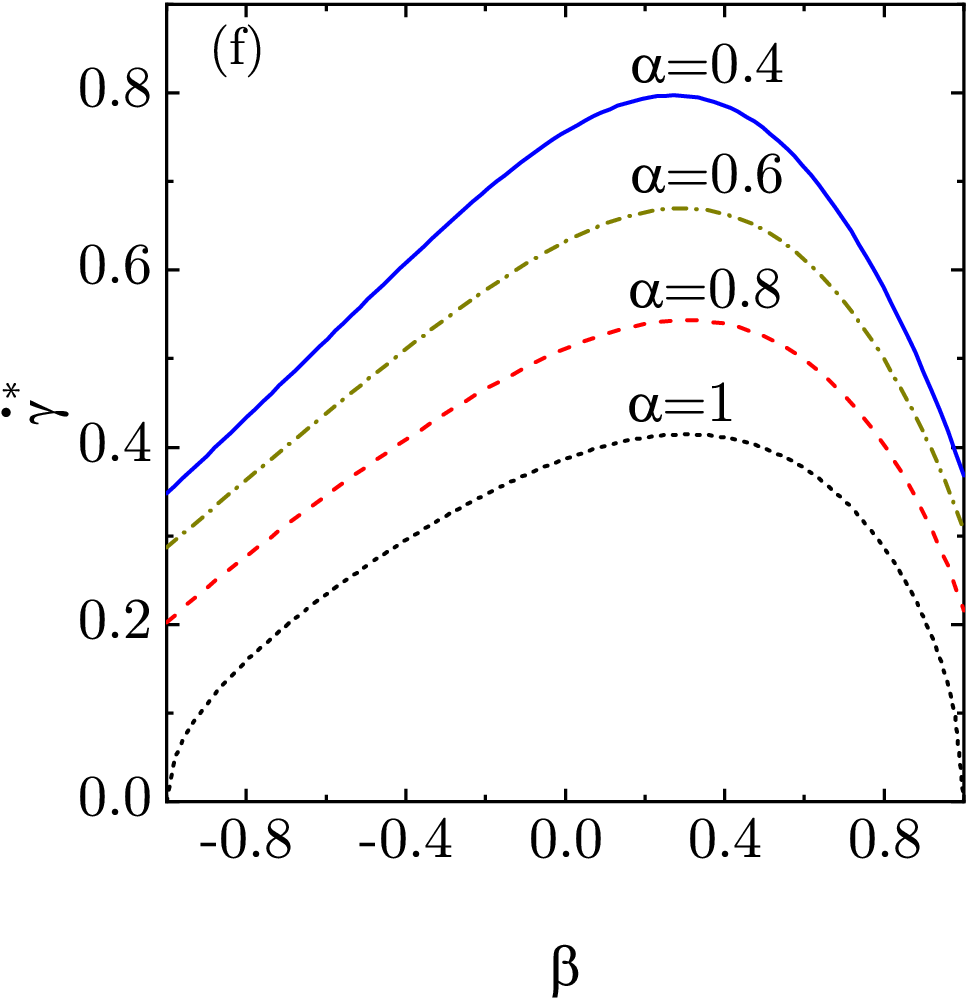}
\caption{
Top panels: reduced normal stress $\Pi_{xx}^*$ (a), reduced shear stress $-\Pi_{xy}^*$ (b), and reduced shear rate $\dot{\gamma}^*$ (c) as functions of the normal restitution coefficient $\alpha$ and the tangential restitution coefficient $\beta$ for $\kappa=\frac{2}{5}$. The displayed planar surfaces correspond to $\alpha=0.4$, $0.6$, $0.8$, and $1$. Bottom panels (d--f): corresponding cross-sections at fixed $\alpha$.}\label{fig2}
\end{figure}

\begin{figure}[tbp]
\centering
\includegraphics[height=0.32\textwidth]{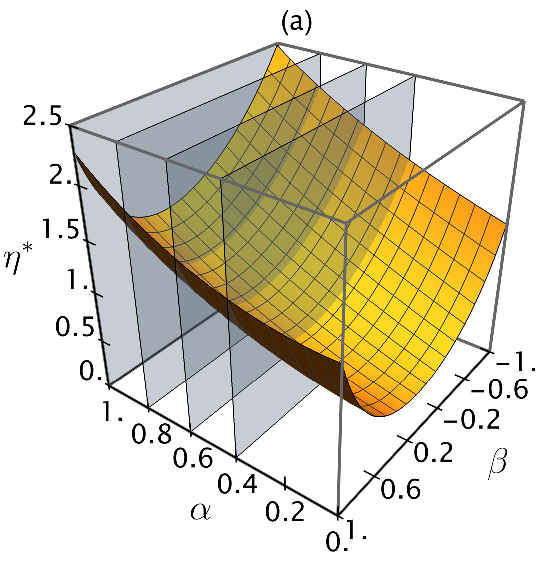}\includegraphics[height=0.32\textwidth]{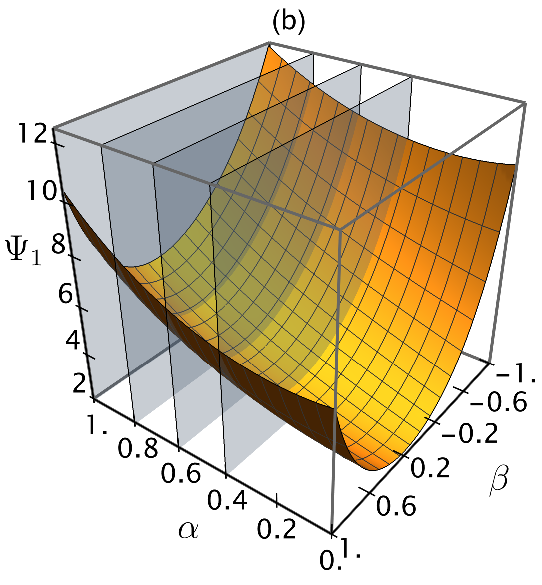}\includegraphics[height=0.32\textwidth]{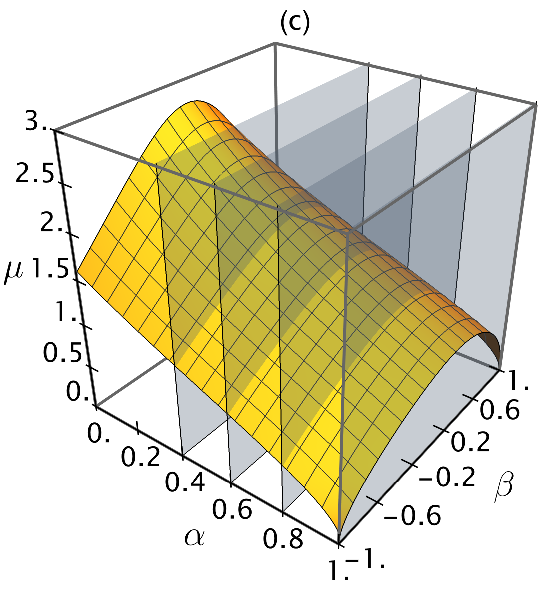}\\
\includegraphics[height=0.28\textwidth]{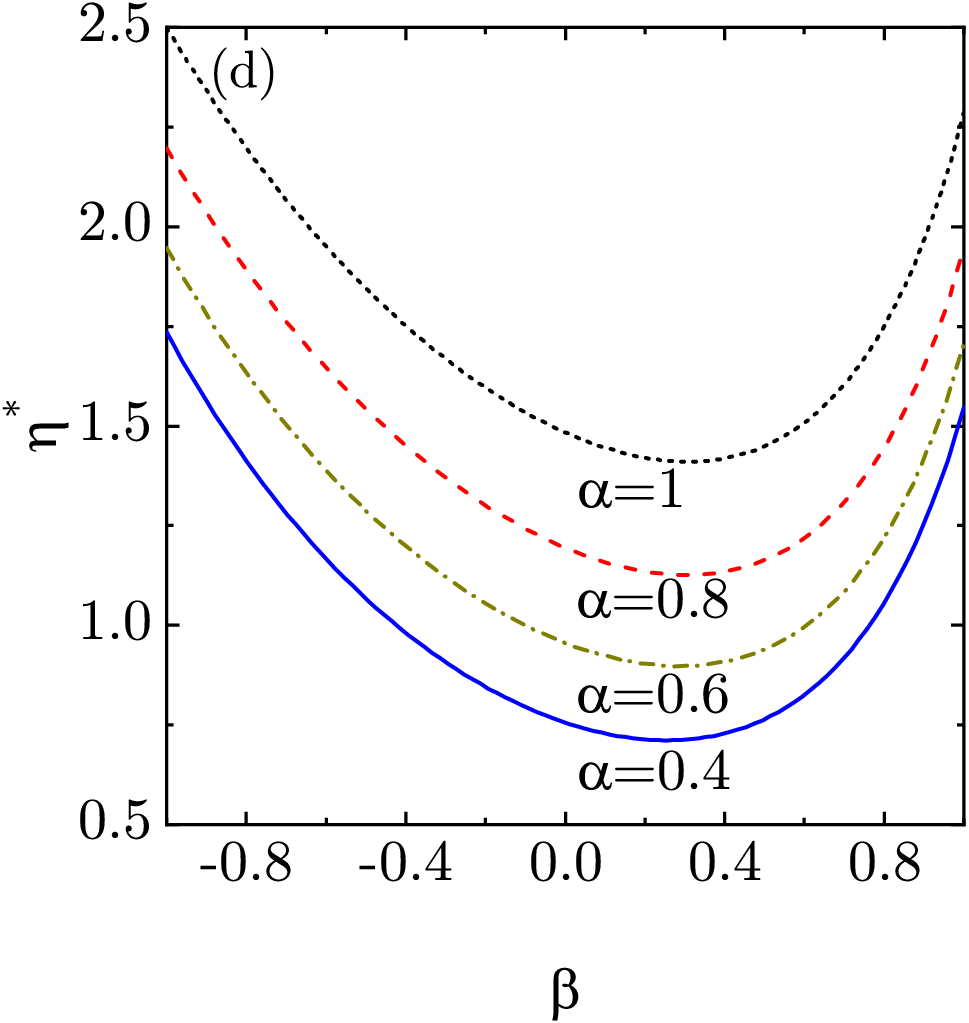}\hspace{0.5cm}\includegraphics[height=0.28\textwidth]{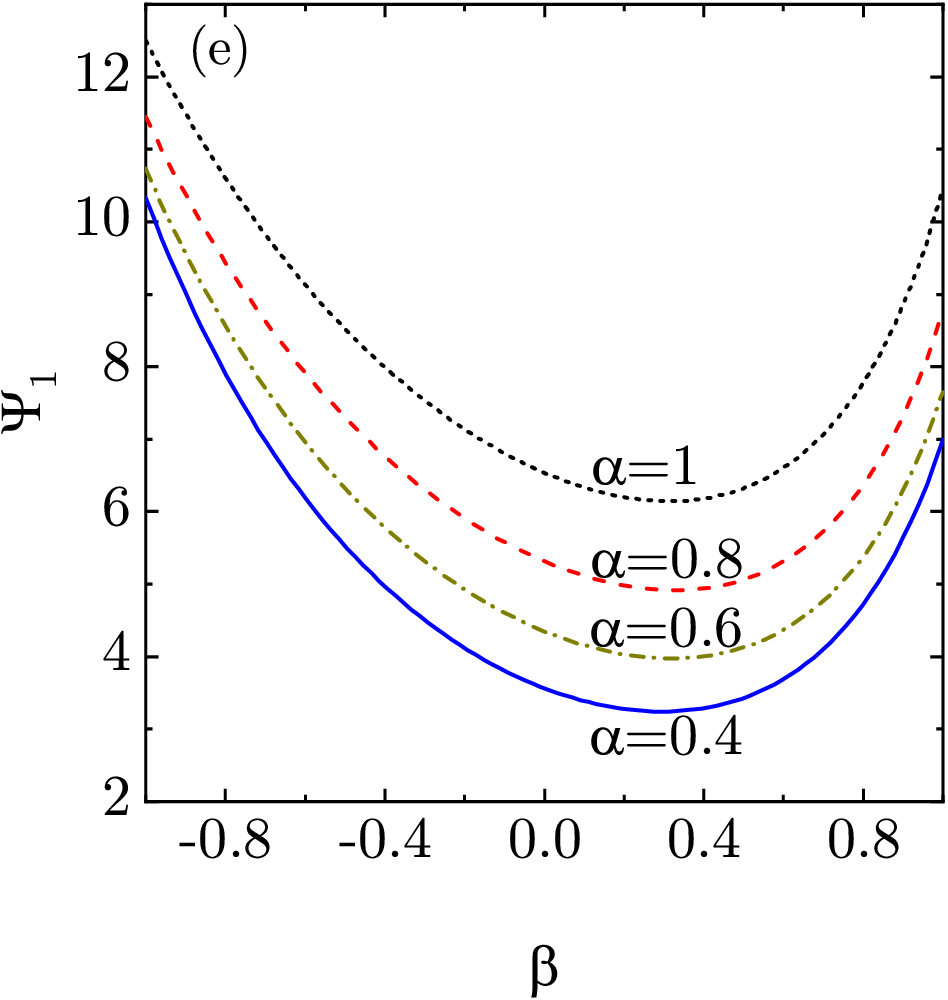}\hspace{0.5cm}\includegraphics[height=0.28\textwidth]{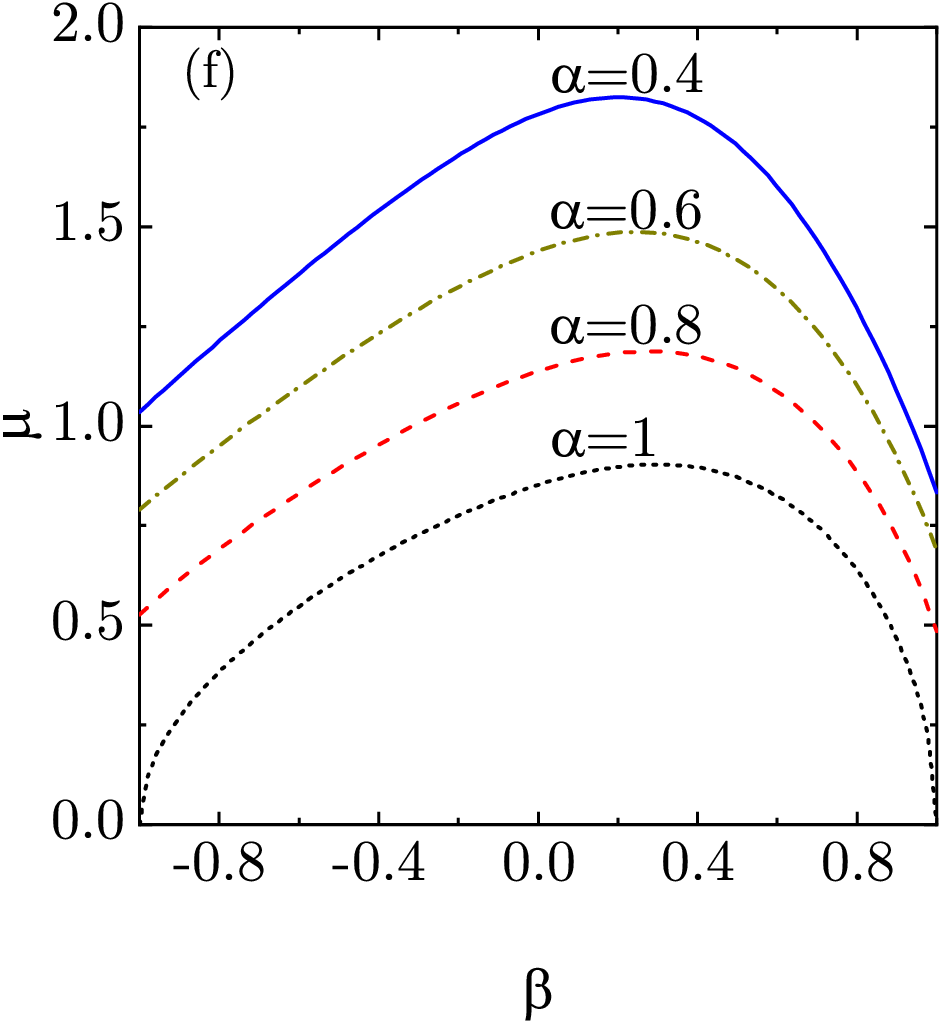}
\caption{
Top panels: reduced shear viscosity $\eta^*$ (a), first viscometric function $\Psi_{1}$ (b), and  friction coefficient $\mu$ (c) as functions of the normal restitution coefficient $\alpha$ and the tangential restitution coefficient $\beta$ for $\kappa=\frac{2}{5}$. The displayed planar surfaces correspond to $\alpha=0.4$, $0.6$, $0.8$, and $1$. Bottom panels (d--f): corresponding cross-sections at fixed $\alpha$.}\label{fig3}
\end{figure}

Figure~\ref{fig2} shows the dependence of $\Pi_{xx}^*$, $-\Pi_{xy}^*$, and $\dot{\gamma}^*$ on $\alpha$ and $\beta$ for uniform spheres, $\kappa=\frac{2}{5}$. At fixed $\alpha$, both the reduced normal stress $\Pi_{xx}^*$ and the reduced shear rate $\dot{\gamma}^*$ exhibit a nonmonotonic dependence on $\beta$, with a maximum around $\beta\sim 0$--$0.4$. At fixed $\beta$, $\Pi_{xx}^*$ and $\dot{\gamma}^*$ increase monotonically as $\alpha$ decreases. The behavior of $\Pi_{xy}^*$ is more intricate due to the combined dependence through $\chi$ and $\psi$.

To further characterize the rheology of the USF, we introduce the reduced non-Newtonian shear viscosity $\eta^*$, the first viscometric function $\Psi_1$, and the friction coefficient $\mu$~\cite{C89}, defined as
\begin{subequations}
\beq
\label{14}
\eta^*\equiv-\frac{\Pi_{xy}^*}{\gda}=\frac{1-\chi/\psi}{\psi}=\frac{2-\Pi_{xx}^*}{2\psi},
\eeq
\beq
\Psi_1\equiv \frac{\Pi_{xx}^*-\Pi_{yy}^*}{\dot{\gamma}^{*2}}=2\frac{1-\chi/\psi}{\psi^2}=\frac{2-\Pi_{xx}^*}{\psi^2},
\eeq
\beq
\label{21c}
\mu\equiv-\frac{\Pi_{xy}^*}{1+\Pi_{yy}^*}=\sqrt{\frac{3}{2}\frac{\chi/\psi}{1-{\chi}/{\psi}}}=\sqrt{\frac{3}{2}\frac{\Pi_{xx}^*}{2-\Pi_{xx}^*}}.
\eeq
\end{subequations}
These quantities are plotted in Fig.~\ref{fig3} for $\kappa=\frac{2}{5}$. The shear viscosity and the first viscometric function increase with increasing $\alpha$ and display a nonmonotonic dependence on $\beta$, with a minimum in the region $\beta\sim 0$--$0.4$. The friction coefficient exhibits a qualitative dependence on $\alpha$ and $\beta$ similar to that of the reduced shear rate.

The relatively large values of the first viscometric function highlight the strong non-Newtonian character of the USF. Moreover, the qualitative dependence of the USF shear viscosity on the coefficients of restitution is opposite to that of the Newtonian shear viscosity (compare Fig.~\ref{fig3}(d) with Fig.~1(a) of Ref.~\cite{SK24}).

In the special case of inelastic and perfectly smooth particles ($\alpha<1$, $\beta=-1$), the above expressions reduce to
\begin{subequations}
\beq
\Pi_{xx}^*=5\frac{1-\al}{4-\al},\quad
\Pi_{xy}^*=-\frac{3}{2}\sqrt{\frac{5}{2}}\frac{\sqrt{1-\al^2}}{4-\al},\quad
\gda=\sqrt{\frac{2}{5}}\frac{4-\al}{6}\sqrt{1-\al^2},
\eeq
\beq
\eta^*=\frac{45}{2(4-\al)^2},\quad \Psi_1=\frac{675}{(4-\al)^3(1+\al)},\quad \mu=\sqrt{\frac{5}{2}\frac{1-\al}{1+\al}}.
\eeq
\end{subequations}
These results agree with those previously obtained for the IMM~\cite{G03a,SG07}.

In the opposite limit of elastic and perfectly rough particles ($\alpha=1$, $\beta=1$), one has $\Pi_{xx}^*=\Pi_{xy}^*=\dot{\gamma}^*=\mu=0$ and
\beq
\eta^*=\frac{5}{2}(1+\ka)^2\frac{3+10\ka}{(1+5\ka)(3+5\ka)},\quad \Psi_1=2\eta^{*2}.
\eeq
Since $\dot{\gamma}^*\to0$ in this limit, the system approaches the
Newtonian regime and the nonlinear shear viscosity reduces to the Newtonian shear viscosity~\cite{SK24} of the Pidduck gas~\cite{P22,CLD65,MSD66,CC70,K10a}. Analogously,  the viscometric function $\Psi_1$ becomes a Burnett coefficient~\cite{K10a}.

\section{Conclusions}
\label{sec5}

In this work we have derived exact results for the USF of a granular gas composed of inelastic and rough Maxwell particles. This study completes the program initiated in our previous papers~\cite{KS22,SK24}, where the collisional moments and the NSF transport coefficients of the IRMM were obtained exactly.

The mean-field character of the IRMM makes it possible to evaluate in closed form the collisional production rates of the second-degree moments and, consequently, to solve exactly the nonlinear rheological problem posed by the steady USF. As in the smooth Maxwell case, the structure of the moment equations allows for an exact determination of the reduced stress tensor and the reduced shear rate without recourse to approximate closures.

A number of remarkable features emerge. First, the rotational-to-translational temperature ratio $\theta=T_r/T_t$ is independent of the coefficient of normal restitution $\alpha$ and depends only on the tangential restitution $\beta$ and the reduced moment of inertia $\kappa$. Moreover, the spin-spin tensor is strictly proportional to the stress tensor, with a proportionality factor that is also independent of $\alpha$. Second, the nonlinear rheological functions---reduced normal stresses, reduced shear stress, shear rate, shear viscosity, viscometric function, and friction coefficient---are obtained in explicit form in terms of two effective parameters, $\chi$ and $\psi$, which generalize the cooling rate and the stress relaxation rate of the smooth model.

The dependence of the rheological properties on $\alpha$ and $\beta$ is highly nontrivial. For uniform spheres ($\kappa=2/5$), the reduced normal stress and the reduced shear rate exhibit a nonmonotonic dependence on $\beta$, with extrema at intermediate roughness, while they increase monotonically with increasing inelasticity (decreasing $\alpha$). The first viscometric function attains relatively large values, clearly illustrating the strongly non-Newtonian character of the USF in the IRMM. Interestingly, the qualitative dependence of the nonlinear shear viscosity on the coefficients of restitution is opposite to that of the Newtonian shear viscosity derived previously.

In the limiting cases of perfectly smooth particles and of elastic perfectly rough particles, our results consistently reduce to the known solutions for the IMM and for the Pidduck IRMM gas, respectively. These reductions provide stringent consistency checks of the results.

Overall, the IRMM provides a rare example of a granular model where non-Newtonian rheology in a genuinely nonlinear state can be analyzed exactly in the presence of both inelasticity and roughness. The results reported here may serve as a benchmark for approximate kinetic theories and for numerical simulations of rough granular gases under shear.

\backmatter

\bmhead{Acknowledgments}

A.S.\ acknowledges financial support from Grant No.~PID2024-156352NB-I00 funded by MCIU/AEI/10.13039/501100011033 and by ERDF/EU, and from Grant No.~GR24022 funded by the Junta de
Extremadura (Spain).

\bmhead{Data availability}
The datasets employed to generate Figs.~\ref{fig1}--\ref{fig3}  are available from the corresponding author on reasonable request.

\section*{Declarations}
\bmhead{Conflict of interest}
The authors declare no conflicts of interest that are relevant to the content of this article.







\end{document}